\documentclass[prodmode,acmtoms]{acmsmall}
\usepackage{amssymb}

\acmVolume{}
\acmNumber{}
\acmArticle{0}
\acmYear{0}
\acmMonth{0}

 \newcommand{\be}{\begin{equation}}
 \newcommand{\ee}{\end{equation}}
 \newcommand{\ea}{\end{array}}

\def\dsp{\displaystyle}
\def\Frac#1#2{\frac
{
 {\raise.6ex
 \hbox{$\displaystyle#1$}}
}
{
 {\lower.6ex
 \hbox{$\displaystyle#2$}}
 }
}

\def\begeq{\begin{equation}
\begin{array}{ll}
}
\def\endeq{\end{array}
\end{equation}}
\usepackage{epsfig}
\usepackage{amssymb}

\def\lneq{{{
{\lower .6ex\hbox{$\scriptscriptstyle<$}}}\atop{\raise.6ex\hbox{$\scriptscriptstyle\sim$}}}}

\def\erfc{{\rm erfc}}

\def\calL{{{\cal L}}}

\def\wt{\widetilde}
\def\tfrac#1#2{{{\lower.6ex
\hbox{$\scriptstyle#1$}}\over 
{\raise.7ex
\hbox{$\scriptstyle#2$}}}}

\def\erfc{{\rm erfc}}

\def\tfrac#1#2{{{\lower.6ex
\hbox{$\scriptstyle#1$}}\over 
{\raise.7ex
\hbox{$\scriptstyle#2$}}}}
\def\eqref#1{(\ref{#1})}
\def\sign{{\rm sign}}
\def\dsp#1{\displaystyle{#1}}

\begin{document}

 \title{Computation of the Marcum Q-function}
 \author{
         AMPARO GIL\\
         Departamento de Matem\'atica Aplicada y Ciencias de la Computaci\'on,
         U. de Cantabria,  39005-Santander, Spain\\   
     JAVIER SEGURA\\
         Departamento de Matem\'aticas, Estad\'{\i}stica y Computaci\'on,
         U. de Cantabria,  39005-Santander, Spain\\   
         \and
      NICO M. TEMME\\
         IAA, 1391 VD 18, Abcoude, The Netherlands
     }

\begin{abstract}

Methods and an algorithm for computing the generalized Marcum $Q-$function ($Q_{\mu}(x,y)$) and the 
complementary function ($P_{\mu}(x,y)$) are described. These functions appear in
problems of  different technical and scientific areas such as, for example, radar detection and communications,
statistics and probability theory, where they are called the non-central chi-square or the non central gamma 
cumulative distribution functions.

The algorithm for computing the Marcum functions
combines different methods of evaluation in different regions: series expansions, integral representations,  
asymptotic expansions, and use of three-term homogeneous recurrence relations. A relative accuracy close to
$10^{-12}$ can be obtained in the parameter region $(x,y,\mu) \in [0,\,A]
\times [0,\,A]\times [1,\,A]$, $A=200$, while for larger parameters the accuracy decreases (close to 
$10^{-11}$ for $A=1000$ and close to $5\times 10^{-11}$ for $A=10000$).
\end{abstract}


\acmformat{Gil, A., Segura, J., Temme, N.M. 2012. Computation of the Marcum Q-function.}

\begin{bottomstuff}
The authors acknowledge financial support from 
{\emph{Ministerio de Econom\'{\i}a y Competitividad}}, projects MTM2009-11686 and MTM2012-34787.

Author's addresses: A. Gil (amparo.gil@unican.es), Departamento de Matem\'atica Aplicada y Ciencias de la Computaci\'on,
         U. de Cantabria,  39005-Santander, Spain; J. Segura (javier.segura@unican.es),
Departamento de Matem\'aticas, Estad\'{\i}stica y Computaci\'on,
         U. de Cantabria,  39005-Santander, Spain;
N.M. Temme (nico.temme@cwi.nl), IAA, 1391 VD 18, Abcoude, The Netherlands. Former address: CWI, Science Park 123, 1098 XG Amsterdam, The Netherlands.
\end{bottomstuff}

\maketitle

\section{Introduction}

We define the generalized Marcum $Q-$function by using the integral representation
\begin{equation}
\label{eq:defQmu}
Q_{\mu} (x,y)=\displaystyle x^{\frac12 (1-\mu)} \int_y^{+\infty} t^{\frac12 (\mu -1)} e^{-t-x} I_{\mu -1} \left(2\sqrt{xt}\right) \,dt,
\end{equation}
where $x\ge0$, $y\ge0$, $\mu >0$, and $I_\mu(z)$ is the modified Bessel function. 
We also use the complementary function
\begin{equation}
\label{eq:defPmu}
P_{\mu} (x,y)=\displaystyle x^{\frac12 (1-\mu)}\int_0^{y} t^{\frac12 (\mu -1)} e^{-t-x} I_{\mu -1} \left(2\sqrt{xt}\right) \,dt,
\end{equation}
and the complementary relation reads
\begin{equation}\label{eq:PQcompl}
P_{\mu}(x,y)+Q_{\mu} (x,y)=1.
\end{equation}

There are other notations for the generalized Marcum function in the literature. Among them, probably the most
 popular one is the following
\begin{equation}
\label{alphabeta}
\widetilde{Q}_{\mu} (\alpha,\beta)=\alpha^{1-\mu}\int_\beta^{+\infty}t^{\mu} e^{-(t^2+\alpha^2)/2} I_{\nu -1}(\alpha t) dt,
\end{equation}
where we have added a tilde in the definition to distinguish it from the definition we are using (\ref{eq:defQmu}). For
$\mu=1$ this coincides with the original definition of the Marcum $Q-$function \cite{Marcum:1960:AST}. This is the notation
used, for instance, in the MATLAB built-in function {\bf marcumq}. The relation with the notation we use is simple:
\begin{equation}
\label{notations}
Q_{\mu}(x,y)=\widetilde{Q}_{\mu} (\sqrt{2x},\sqrt{2y}),
\end{equation}
and similarly for the $P-$function.

The generalized Marcum $Q-$function is an important function used in radar detection and communications, see
\cite{Marcum:1960:AST,Rice:1968:UAE,Robertson:1976:CON}. In this field, $\mu$ is the number of
independent samples of the output of a square-law detector. In
our analysis $\mu$ is not necessarily a positive integer number and we will consider real values $\mu \ge 1$. 

These functions also occur in statistics and probability theory, where they are called non-central chi-square 
(semi-integer $\mu$) or non-central 
gamma cumulative distributions 
\cite{Ashour:1990:OTC,Cohen:1988:CAT,Dyrting:2004:ENC,Knusel:1996:CNG,Robertson:1976:CON,Ross:1999:ACN}. The
 central gamma cumulative distribution is in fact the incomplete gamma function, and the relation of the Marcum functions to
 the incomplete gamma functions is given in \S\ref{sec:serexp}.

In this paper we describe methods and an algorithm for computing the functions $P_{\mu}(x,y)$ and $Q_{\mu} (x,y)$ for a
 large range of the  parameters $\mu, x, y$.  We consider series expansions in terms of the incomplete gamma functions,
 recurrence relations, asymptotic expansions, and numerical quadrature. 
A Fortran 90 module implementing the algorithm is tested,
and we conclude it provides a relative accuracy close to $\sim 10^{-12}$ in the parameter 
region $(x,y,\mu) \in [0,\,A] \times [0,\,A] \times [1,\,A]$, $A=200$, 
while for larger parameters the accuracy decreases (close to 
$10^{-11}$ for $A=1000$ and better than $10^{-10}$ for $A=10000$).

We include comparisons with the MATLAB built-in function {\bf marcumq}; this function is 
restricted to integer positive values of the parameter $\mu$, differently from our 
Fortran 90 module which allows the computation of the Marcum functions for real values 
(greater or equal to 1) of that parameter. Also, the MATLAB function computes the Marcum 
$Q-$function but not the complementary $P-$function: computing this function simply 
as $1-Q$ when $Q$ is close to 1 can lead to serious cancellation problems. Our tests reveal some bugs 
in certain parameter regions when computing the Marcum $Q-$function using MATLAB.

\section{Properties of the Marcum functions}\label{sec:props}

\subsection{Series expansions}\label{sec:serexp}

Substituting the Maclaurin series for the modified Bessel function 
\begin{equation}\label{eq:defImu}
I_\mu(z)=\left(\tfrac12z\right)^\mu\sum_{n=0}^\infty \frac{\left(\frac14z^2\right)^n}{n!\,\Gamma(\mu+n+1)}
\end{equation}
into the integral representations \eqref{eq:defQmu} and \eqref{eq:defPmu}, we readily obtain the series expansions
\begin{equation}\label{qratio}
\begin{array}{l}
\dsp{P_{\mu} (x,y)=e^{-x}\sum_{n=0}^{\infty} \frac{x^n}{n!}  P_{\mu +n}(y),}\\[8pt]
\dsp{Q_{\mu} (x,y)=e^{-x}\sum_{n=0}^{\infty} \frac{x^n}{n!}  Q_{\mu +n}(y).}
\end{array}
\end{equation}

These expansions are in terms of the incomplete gamma function ratios defined by
\begin{equation}
\label{eq:defincgam}
P_{\mu} (x)=\frac{\gamma (\mu,x)}{\Gamma(\mu)},\quad
Q_{\mu} (x)=\frac{\Gamma (\mu,x)}{\Gamma(\mu)},
\end{equation}
where  the standard incomplete gamma functions are defined by
\begin{equation}\label{eq:igfs}
\gamma (\mu,x)=\int_{0}^x t^{\mu-1} e^{-t} \,dt,\quad
\Gamma (\mu,x)=\int_{x}^{+\infty} t^{\mu-1} e^{-t} \,dt, \quad \Re\mu>0.
\end{equation}
Again, we have the complementary relation $P_\mu(x)+Q_\mu(x)=1$. In \cite{Gil:2012:IGR}, 
algorithms are given for the computation of the incomplete gamma function ratios $P_{\mu} (y)$ and $Q_{\mu} (y)$
appearing in (\ref{qratio}).

From these relations we obtain the particular values
\begin{equation}\label{qmval}
\begin{array}{ll}
Q_{\mu} (x, 0)=1,&Q_{\mu}(x ,+\infty)=0,\\[8pt]
Q_{\mu} (0,y)=Q_\mu(y),\quad& Q_{\mu}(+\infty,y)=1,\\[8pt]
Q_{+\infty} (x ,y )=1,&
\end{array}
\end{equation}
and similar complementary relations for $P_\mu(x,y)$.

In the numerical algorithm we compute both functions $P_{\mu} (x,y)$ and $Q_{\mu} (x,y)$. The {\em primary function}  in the algorithm is the smallest one of $P_{\mu} (x,y)$ and $Q_{\mu} (x,y)$, and the primary function will be computed first. The other one follows from the complementary relation in \eqref{eq:PQcompl}.

As will follow from the relations given later (and in less detail from the relations in \eqref{qmval}), the transition in the $(x,y)$ quarter plane from small values of $Q_{\mu} (x,y)$ to values close to unity occurs for large values of $\mu, x,y $ across the  line $y=x+\mu$, and above this line in the $(x,y)$ quarter plane $Q_{\mu} (x,y)$ is taken as the primary function. Below this   line the complementary function $P_{\mu} (x,y)$ is taken as the primary function.

\subsection{Recurrence relations}

Considering integration by parts in the integrals in \eqref{eq:defQmu} and  \eqref{eq:defPmu}, together with the relation $z^\mu I_{\mu-1}(z)=\frac{d}{dz}\left(z^\mu I_\mu(z)\right)$, it
is easy to see that the Marcum functions satisfy the following first order difference equations (see, for instance, \cite{Temme:1993:ANA}
\footnote{We note that a factor $e^{-y}$ is missing in \cite[Eq.~(1.4)]{Temme:1993:ANA}.}):

\begin{equation}
\label{RecQ}
\begin{array}{ll}
\dsp{Q_{\mu+1}(x,y)=Q_{\mu} (x,y)+\left(\Frac{y}{x}\right)^{\mu/2} e^{-x-y} I_{\mu} (2\sqrt{xy})},\\[8pt]
\dsp{P_{\mu+1}(x,y)=P_{\mu} (x,y)-\left(\Frac{y}{x}\right)^{\mu/2} e^{-x-y} I_{\mu} (2\sqrt{xy})}.
\end{array}\end{equation}

It is possible to eliminate the Bessel function appearing in (\ref{RecQ}) and to obtain a homogeneous third order recurrence relation \cite{Temme:1993:ANA}:

\begin{equation}
\label{RecQ2}
xQ_{\mu+2}(x,y)=(x-\mu)Q_{\mu+1} (x,y)+ (y+\mu)Q_{\mu} (x,y) 
-yQ_{\mu-1} (x,y).
\end{equation}  
The function $P_\mu(x,y)$  and every constant (with respect to $\mu$) satisfy the same relation. These recurrence relations can be useful for testing.

This first order inhomogeneous equation in \eqref{RecQ} can be used for computing the Marcum $Q-$function in the direction of increasing $\mu$. Observe that the right-hand side in the relation for $Q_\mu(x,y)$ has only positive terms. Also, considering
Perron's theorem for this difference equation (\cite[Thm. 4.17]{Gil:2007:NSF}), it is easy to check that the recurrence relation admits a minimal
(or recessive) solution and that the dominant solutions $y_\mu$ are such that
\begin{equation}
\lim_{\mu\rightarrow +\infty}\Frac{y_{\mu+1}}{y_\mu}=1 ,
\end{equation}
as corresponds to the Marcum $Q-$function, which is therefore dominant and can be computed by forward recursion.

When we consider Perron's theorem, which is also known as  the Perron-Kreuser theorem, for the homogeneous recursion in \eqref{RecQ2},
 it follows that the Marcum $Q-$function is neither
minimal nor dominant and therefore it cannot be computed neither in the forward or backward direction. A possible way out 
is to combine two first order recursions of \eqref{RecQ} to obtain the three-term homogeneous recurrence relation

\begin{equation}
\label{TTRR}
y_{\mu+1} -(1+c_\mu) y_\mu + c_\mu y_{\mu-1}=0,\quad c_\mu=\sqrt{\frac{y}{x}}\Frac{I_\mu \left(2\sqrt{xy}\right)}{I_{\mu-1}\left(2\sqrt{xy}\right)}.
\end{equation}

Both $Q_\mu (x ,y)$ and the complementary function $P_\mu (x ,y)$ satisfy (\ref{TTRR}). $Q_\mu(x,y)$ is dominant and can be computed in the forward recursion, while $P_\mu(x,y)$ is minimal and has to be computed in the backward direction.

An advantage of the relation in \eqref{TTRR} is that ratios of Bessel functions appear, which are slowly varying with respect to $\mu$.   Because dominant factors in the Bessel function representations  disappear in the ratios, these are
free of overflow/underflow problems. In addition, for the ratios of Bessel functions continued fraction representations
can be used.

It is interesting to observe that Pincherle's theorem \cite[Thm. 2.7]{Gil:2007:NSF} allows us to compute a ratio of $P_\mu (x ,y)$ as a continued
fraction,  the coefficients of which can be represented as continued fractions themselves, namely

\begin{equation}
\label{cfP}
\Frac{P_\mu (x ,y)}{P_{\mu-1}(x,y )}=\Frac{c_\mu}{1+c_\mu\,-}\,\Frac{c_{\mu+1}}{1+c_{\mu+1}\,-}\,\ldots\,.
\end{equation}

\subsection{Derivatives and monotonicity}

 Taking the derivative with respect to $y$ in (\ref{eq:defQmu}) and using (\ref{RecQ}) we have
\begin{equation}
\Frac{\partial Q_\mu(x,y)}{\partial y}=Q_{\mu-1}(x,y)-Q_{\mu}(x,y) .
\end{equation}

Taking the derivative with respect to $x$, and using the relation $I_{\nu}'(z)=I_{\nu+1}(z)+\frac{\nu}{z}I_{\nu}(z)$, we obtain
\begin{equation}
\Frac{\partial Q_\mu(x,y)}{\partial x}=Q_{\mu+1}(x,y)-Q_{\mu}(x,y) .
\end{equation}

And using (\ref{RecQ}) we see that $Q_\mu (x,y)$ ($P_\mu (x,y)$) is an increasing (decreasing) 
function of $x$ and a decreasing (increasing) function of $y$. With respect to $\mu$, $Q_\mu(x,y)$ is increasing and $P_\mu(x,y)$ is decreasing.

\section{Using the series expansions}\label{sec:userexp}
The series representations in \eqref{qratio} can be computed by using the algorithms for the incomplete gamma ratios described in  \cite{Gil:2012:IGR}. The recurrence relations
\begin{equation}\label{eq:serexp01}
\begin{array}{ll}
\dsp{Q_{\mu+1}(y)=Q_\mu(y)+\frac{y^\mu e^{-y}}{\Gamma(\mu+1)},}\\[8pt]
\dsp{P_{\mu+1}(y)=P_\mu(y)-\frac{y^\mu e^{-y}}{\Gamma(\mu+1)},}
\end{array}
\end{equation}
are stable for $Q_\mu(y)$ in the forward direction, and for $P_\mu(y)$ in the backward direction. 
For the series representation of $P_\mu(x,y)$ we need to find a truncation number $n_0$ of the infinite series, 
compute $P_{\mu+n_0}\left(y\right)$  and perform the summation and recursion in \eqref{qratio} with $n=n_0, n_0-1, \ldots,0$.

The recursion for $P_{\mu+n}\left(y\right)$ is written in the backward form
\begin{equation}\label{eq:serexp02}
P_{\mu+n}(y)=P_{\mu+n+1}(y)+\frac{y^{\mu+n} e^{-y}}{\Gamma(\mu+n+1)},
\end{equation}
and we write the series for  $P_\mu(x,y)$ in the form
\begin{equation}\label{eq:serexp03}
P_{\mu} (x,y)\simeq e^{-x}P_\mu(y)\sum_{n=0}^{n_0} \frac{x^n}{n!}  \frac{P_{\mu +n}(y)}{P_{\mu}(y)},
\end{equation}
where we assume that the first neglected term is smaller than a prescribed precision~$\varepsilon$. Observe that 
the value of the first term in the sum is $1$, and then we expect that by considering $S_{n_0}<\varepsilon$, with $S_{n_0}$
the first neglected term, the sum will reach a relative accuracy better than $\varepsilon$.

The first neglected term 
\begin{equation}
S_{n_0}=\frac{x^{n_0+1}}{(n_0+1)!}  \frac{P_{\mu +n_0+1}(y)}{P_{\mu}(y)}
\end{equation} 
can be bounded using some monotonicity properties of the incomplete gamma function. Using (\ref{eq:defincgam}) we have
\begin{equation}
S_{n_0}=\frac{x^{n_0+1}}{(n_0+1)!(\mu)_{n_0+1}}\frac{\gamma (\mu+1,y)}{\gamma (\mu,y)}\frac{\gamma (\mu+2,y)}{\gamma (\mu+1,y)}
\ldots \frac{\gamma (\mu+n_0+1,y)}{\gamma (\mu+n_0,y)}.
\end{equation}
Now, it is easy to check that $h_{\nu}(y)=\frac{\gamma (\nu,y)}{\gamma (\nu-1,y)}<y$. Indeed, it is known that $h_{\nu}(y)$
is monotonically increasing as a function on $\nu$ \cite{Qi:2002:MRI} and, on the other hand, it is easy to check that 
$\lim_{\nu\rightarrow +\infty} h_{\nu}(y)=y$ (for instance, by using 
\cite[8.11.4]{Paris:2010:IGR}); these two facts imply that $h_{\nu}(y)<y$. With this information, we conclude that
\begin{equation}
S_{n_0}<\frac{x^{n_0+1}y^{n_0+1}}{(n_0+1)!(\mu)_{n_0+1}},
\end{equation}
and taking 
\begin{equation}
\frac{(x y)^{n_0+1}(\mu-1)!}{(n_0+1)!(\mu+n_0)!}=\varepsilon,
\end{equation}
it is guaranteed that the first neglected term is smaller than $\varepsilon$. Now we use the Stirling approximation for
the factorials in the denominators, which is a lower bound for the factorials. This translates into finding the
solution of $f(n)=0$ for
\begin{equation}
f(n)=(n+\mu+1/2)\ln (n+\mu)+(n-1/2)\ln n -2n-n \ln (xy)-C,\,
C=\ln\left(\Frac{\Gamma (\mu)}{2\pi\varepsilon}\right)+\mu.
\end{equation}
If $f(n^*)=0$ with $n^*$ the larger solution then it is guaranteed that $n_0=n^*-1$ is such that $S_{n_0}<\varepsilon$.
 
The analysis becomes more simple considering some tiny modifications and taking 
\begin{equation}
f(n)=(n+\mu)\ln (n+\mu)+n\ln n -2n -n \ln (xy) -C .
\end{equation}
For this function the derivative with respect to $n$ is given by
 \begin{equation}\label{eq:serexp07}
f^\prime(n)=\ln\frac{n(n+\mu)}{xy},
 \end{equation}
and this first derivative vanishes if $n=n_e=(-\mu+\sqrt{\mu^2+4xy})/2$, which is a minimum because $f^{(2)}(n)>0$.
Starting at the right of this minimum we can use Newton's method to solve $f(n)=0$; 
convergence will be certain because $f^\prime(n)>0$ at the right of the minimum and $f^{(2)}(n)>0$.

We observe that for small values of $P_{\mu}(x,y)$ (close to the underflow limit), it may happen that the last
terms in the series can not be computed because the incomplete gamma ratios may underflow. Similarly, in the the series
for the $Q_{\mu}(x,y)$ it may happen that the first values of $Q_{\mu+n}(y)$ in (\ref{qratio}) may underflow, but
not the total sum. This means that some
error degradation can be expected for values close to underflow. We will discuss this issue later in 
\S\ref{algorithm}.

\section{Asymptotic representations}\label{sec:asrep}
In \cite{Temme:1993:ANA} two types of asymptotic expansions have been derived: one with  small values of $\mu$ and one with large values.
 In both cases the expansion is given in terms of the
 complementary error function
\begin{equation}\label{eq:ar01}
\erfc\,z =\frac{2}{\sqrt{\pi}}\int_z^\infty e^{-t^2}\,dt;
\end{equation}
in this definition $z$ may assume any finite (complex) value.

\subsection{Asymptotic representations for large  \protect{\boldmath{$xy$}}}\label{sec:asrepmus}
We summarize the results given in \cite{Temme:1993:ANA}. We use the function
\begin{equation}\label{eq:mus01}
F_\mu(\xi,\sigma):=\int_\xi^\infty
e^{-(\sigma+1)t}I_\mu(t)\,dt,\quad \sigma>0,
\end{equation}  
and we have
\begin{equation}\label{eq:mus02}
\begin{array}{@{}r@{\;}c@{\;}l@{}}
Q_\mu(x,y)&=&\tfrac12\rho^\mu \left(F_{\mu-1}(\xi,\sigma)-\frac1\rho
F_\mu(\xi,\sigma)\right),\quad y>x,\\[8pt]
P_\mu(x,y)&=&\tfrac12\rho^\mu \left(\frac1\rho
F_\mu(\xi,\sigma)-F_{\mu-1}(\xi,\sigma)\right),\quad y<x,
\end{array}\end{equation}
where the parameters are defined  by
\begin{equation}\label{eq:mus03}
\xi=2\sqrt{{xy}},\quad\sigma=
\frac{(\sqrt{{y}}-\sqrt{{x}})^2}{\xi},\quad\rho=\sqrt{{\frac y x}}.
 \end{equation}    

We assume that $\xi$ is large and substitute 
the well-known expansion\footnote{http://dlmf.nist.gov/10.40.E1}
\begin{equation}\label{eq:mus04}
e^{-t}I_\mu(t)\sim\frac1{\sqrt{{2\pi t}}}\sum_{n=0}^\infty(-1)^n\frac{A_n(\mu)
}{t^n}, \quad A_n(\mu)
=\frac{2^{-n}\Gamma(\tfrac12+\mu+n)}{n!\Gamma(\tfrac12+\mu-n)},\quad n=0,1,2,\ldots,
\end{equation}
into \eqref{eq:mus01}. This gives the expansion 
\begin{equation}\label{eq:mus05}
F_\mu(\xi,\sigma)\sim\frac1{\sqrt{{2\pi}}}\sum_{n=0}^\infty(-1)^nA_n(\mu)
\Phi_n,
\end{equation}
 where 
$\Phi_n$ is an incomplete gamma function (see \eqref{eq:igfs}) 
\begin{equation}\label{eq:mus06}
\Phi_n=\int_\xi^\infty e^{-\sigma
t}t^{-n-\tfrac12}\,dt=\sigma^{n-1/2}\Gamma\left(\tfrac12-n,\sigma\xi\right).
\end{equation}

The function $\Phi_0$ can be written in terms of  the complementary error function (see \eqref{eq:ar01})
\begin{equation}\label{eq:mus07}
\Phi_0=\sqrt{{\pi/\sigma}}\,\erfc\,\sqrt{{\sigma\xi}}=\sqrt{{\pi/\sigma}}\,\erfc(\sqrt{{y}}-\sqrt{{x}}).
\end{equation}
Further terms can be obtained from the recursion
\begin{equation}\label{eq:mus08}
\left(n-\tfrac12\right)\Phi_n=-\sigma\Phi_{n-1}+e^{-\sigma\xi}\xi^{-n+\tfrac12},\quad n=1,2,3,\ldots.\end{equation}

Using \eqref{eq:mus02} and \eqref{eq:mus05} we obtain 
\begin{equation}\label{eq:mus09}
Q_\mu(x,y)\sim\sum_{n=0}^\infty\Psi_n,
\end{equation}
where
\begin{equation}\label{eq:mus10}
\Psi_n=\frac{\rho^\mu}{2\sqrt{{2\pi}}}(-1)^n
\left(A_n(\mu-1)-\frac1\rho A_n(\mu)
\right)\Phi_n.
\end{equation}

For $P_{\mu}(x,y)$ the expansion reads
\begin{equation}\label{eq:mus11}
P_\mu(x,y)\sim \sum_{n=0}^\infty \wt\Psi_n,
\end{equation}
where $\wt\Psi_n=-\Psi_n$, $n\ge1$, and
\begin{equation}\label{eq:mus12}
\wt\Psi_0=\tfrac12\rho^{\mu-\tfrac12}\erfc(\sqrt{{x}}-\sqrt{{y}}). 
\end{equation}

Information on the asymptotic nature and error bounds of expansion \eqref{eq:mus05} can be found
in \cite{Temme:1986:ADI}, where  numerical aspects of recursion \eqref{eq:mus08} are discussed as well.
 Recursion of the functions $\Phi_n$ should begin at some point $n_0$ near $\sigma\xi$, and from $n_0$ backward or forward recursion has to be used. These results are based on \cite{Gautschi:1961:RCC}, where exponential integrals are considered, which are related to the incomplete gamma functions.

The expansions in \eqref{eq:mus09} and \eqref{eq:mus11} hold for large values of $\xi$, uniformly with respect to $\sigma\in[0,\infty)$.   

Note that the integral defining $F_\mu(\xi,\sigma)$ in \eqref{eq:mus01} becomes undetermined when $\sigma=0$.
However, since we use a combination of two $F-$functions in \eqref{eq:mus02}, and $\rho$ tends to unity as
$\sigma\to0$, that is, as $x\to y$, the left-hand sides in \eqref{eq:mus02}  are well defined when $x=y$.

\subsection{Asymptotic representations for large  \protect{\boldmath{$\mu$}}}\label{sec:asrepmul}

For numerical calculations the large $\mu$ expansion given in \cite[Eq.~(4.3)]{Temme:1993:ANA} is not suitable, and in this section we derive an alternative expansion with coefficients that are easier to evaluate than those derived earlier.

We consider both $Q_\mu(x,y)$ and $P_\mu(x,y)$ and use the integral representations
given in (\ref{eq:defQmu}) and (\ref{eq:defPmu}).

For using the asymptotic properties of the Bessel function it is convenient to consider this function with order $\mu$. Furthermore, we scale $x$ and $y$ and write
\begin{equation}\label{eq:asrep03}
Q_{\mu+1} (\mu x,\mu y)=\int_{\mu y}^{\infty} \left(\frac{t}{\mu x}\right)^{\frac12\mu} e^{-t-\mu x} I_{\mu} (2\sqrt{\mu xt}) \,dt.
\end{equation}
A change of variable $2\sqrt{\mu xt}=\mu z$ gives
\begin{equation}\label{eq:asrep04}
Q_{\mu+1} (\mu x,\mu y)=\frac{\mu e^{-\mu x}}{(2x)^{\mu+1}}
\int_{\xi}^{\infty} z^{\mu+1}e^{-\frac{\mu}{4x}z^2} I_{\mu} (\mu z) \,dz,
\end{equation}
\begin{equation}\label{eq:asrep05}
P_{\mu+1} (\mu x,\mu y)=\frac{\mu e^{-\mu x}}{(2x)^{\mu+1}}
\int_0^{\xi} z^{\mu+1}e^{-\frac{\mu}{4x}z^2} I_{\mu} (\mu z) \,dz,
\end{equation}
where
\begin{equation}\label{eq:asrep06}
\xi=2\sqrt{xy}.
\end{equation}

\subsubsection{Asymptotic expansion of the modified Bessel function}\label{sec:asmb}
For deriving an asymptotic expansion of the functions $Q_\mu(x,y)$ and $P_\mu(x,y)$ 
we use the expansion\footnote{http://dlmf.nist.gov/10.41.E3}
\begin{equation}\label{eq:mb01}
I_{\mu}(\mu z)\sim\frac{1}{ \sqrt{2\pi\mu}}\frac{e^{\mu\eta(z)}}{(1+z^2)^{1/4}}
\sum_{k=0}^{\infty}\frac{u_k(t)}{\mu^k}, \quad \mu\to\infty, \quad z\ge0,
\end{equation}
where
\begin{equation}\label{eq:mb02}
t=\frac{1}{\sqrt{1+z^2}}, \qquad
\eta(z)=\sqrt{1+z^2}+\log\frac{z}{1+\sqrt{1+z^2}}.
\end{equation}

The first coefficients $u_k(t)$ are
\begin{equation}\label{eq:mb03}
u_0(t)=1, \qquad u_1(t)=\frac{3t-5t^3}{24}, 
\qquad u_2(t)=\frac{81t^2-462t^4+385t^6}{1152},
\end{equation}
and other coefficients can be obtained by applying the formula
\begin{equation}\label{eq:mb04}
u_{k+1}(t)=\tfrac 12 t^2(1-t^2)u_k'(t)+\tfrac 18 \int_0^t (1-5s^2)u_k(s) \,ds,
\quad k=0,1,2,\ldots.
\end{equation}

\subsection{Asymptotic expansion of \protect{\boldmath{$Q_{\mu+1}(\mu x,\mu y)$}}}\label{sec:asqmu}
We write \eqref{eq:asrep04}  in the form
\begin{equation}\label{eq:qas01}
Q_{\mu+1} (\mu x,\mu y)=\frac{\mu e^{-\mu x}}{(2x)^{\mu+1}}
\int_{\xi}^{\infty} z e^{-\mu\phi(z)}e^{-\mu\eta(z)}I_{\mu} (\mu z) \,dz,
\end{equation}
where
\begin{equation}\label{eq:qas02}
\phi(z)=-\ln  z+\frac{1}{4x}z^2-\eta(z).
\end{equation}

The saddle point follows from the equation $\phi^\prime(z)=0$. We have $\eta^\prime(z)= \sqrt{1+z^2}/z$, and the saddle point  is obtained by solving the equation
\begin{equation}\label{eq:qas03}
\phi^\prime(z)=-\frac{1}{z}+\frac{z}{2x}-\frac{\sqrt{1+z^2}}{z}=\frac{-2x+z^2-2x\sqrt{1+z^2}}{2xz}=0.
\end{equation}
It follows that the positive saddle point $z_0$ is given by
\begin{equation}\label{eq:qas04}
z_0=2\sqrt{x(1+x)}.
\end{equation}

The saddle point is located outside the interval of integration if $\xi>z_0$, that is, when $y>x+1$, in which case  $Q_{\mu+1}(\mu x,\mu y)\lesssim\frac12$ (when the parameters are large). As discussed earlier, see also  \cite[\S4]{Temme:1993:ANA}, the relation  $y=x+1$ (in the scaled variables)  indicates a transition from small values of $Q_{\mu+1}(\mu x,\mu y)$ (when $y>x+1$) to values close to 1 when ($y<x+1$) (see also the first line of \eqref{qmval}).

So, for the $Q-$function the case $\xi>z_0$ is of special interest, because in that case this function is smaller than the $P-$function.  We will derive an expansion in which the cases $y<x+1$, $y=x+1$, and $y>x+1$ are included; however, for $y<x+1$ we use an expansion for the $P-$function, see \S\ref{sec:aspmu}. 

We use  a method introduced in \cite{Bleistein:1966:UAE}, see also \cite[pp.~344--351]{Olver:1997:ASF}.
As a standard form in asymptotic analysis with the above described features (a saddle point coalescing with an end point of integration) we can consider
\begin{equation}\label{eq:qas05}
\int_0^\infty e^{-\mu(\frac12 w^2-\zeta w)} f(w)\,dw,
\end{equation}
with saddle point $\zeta$, which may be any  complex number. We transform the integral in \eqref{eq:qas01} into this form by writing
\begin{equation}\label{eq:qas06}
\phi(z)-\phi(\xi)=\tfrac12w^2-\zeta w,
\end{equation}
where $\zeta$ has to be determined. The point $z=\xi$ corresponds to $w=0$, and we assume that the $z-$saddle point $z_0$ corresponds to the 
$w-$saddle point at $w=\zeta$. Also, we assume that $\sign(z-z_0)=\sign(w-\zeta)$.
This gives
\begin{equation}\label{eq:qas07}
\tfrac12\zeta^2=\phi(\xi)-\phi(z_0).
\end{equation}
Because $z_0$ is the saddle point, the right-hand side is always nonnegative. 

The sign of $\zeta$ is chosen as follows. As  mentioned after \eqref{eq:qas04}, the saddle point 
$z_0$ is inside the domain of integration if $y<x+1$. The same condition is used for the sign of~$\zeta$:
\begin{equation}\label{eq:qas08}
\zeta=\sign(x+1-y)\sqrt{2\left(\phi(\xi)-\phi(z_0)\right)}.
\end{equation}
For the numerical evaluation of $\zeta$ from this relation, we refer to \S\ref{sec:expz}.

The transformation \eqref{eq:qas06} applied to \eqref{eq:qas01} gives
\begin{equation}\label{eq:qas09}
Q_{\mu+1}(\mu x,\mu y)=
\sqrt{\frac{\mu}{2\pi}} e^{-\frac12\mu\zeta^2}
\int_{0}^{\infty} e^{-\mu(\frac12w^2-\zeta w) } f(w)\,dw,
\end{equation}
where 
\begin{equation}\label{eq:qas10}
f(w)=\sqrt{2\pi\mu} \frac{z}{2x} e^{-\mu\eta(z)}I_\mu(\mu z)\frac{dz}{dw}.
\end{equation}

In the representation in \eqref{eq:qas09} we have used the relation
\begin{equation}\label{eq:qas11}
(2x)^{-\mu}e^{-\mu x}e^{-\mu\phi(\xi)}=e^{-\frac12\mu\zeta^2},
\end{equation}
which follows from observing that
\begin{equation}\label{eq:qas12}
\phi(z_0)=-x-\ln(2x).
\end{equation}

In Bleistein's method an asymptotic expansion is obtained by expanding (interpolating, 
actually) this function at the points $w=0$ and $w=\zeta$. We modify this method, as in 
\cite[pp.~346--348]{Olver:1997:ASF}, by expanding at $w=\zeta$ only. Also, we replace the Bessel function with its 
asymptotic expansion in \eqref{eq:mb01}. In this way, we obtain
\begin{equation}\label{eq:qas13}
Q_{\mu+1}(\mu x,\mu y)\sim\sqrt{\frac{\mu}{2\pi}}\sum_{k=0}^\infty \frac{\Phi_k}{\mu^k}, 
\end{equation}
where
\begin{equation}\label{eq:qas14}
\Phi_k=e^{-\frac12\mu\zeta^2}\int_{0}^\infty e^{-\mu(\frac12w^2-\zeta w)} f_k(w)\,dw, 
\end{equation}
with (see \eqref{eq:mb01}) 
\begin{equation}\label{eq:qas15}
f_k(w)=\frac{z}{2x}\frac{u_k(t)}{(1+z^2)^{\frac14}}\frac{dz}{dw},
\end{equation}
and $t$ is as in \eqref{eq:mb02}. The relation between $z$ and $w$ follows from the transformation given in \eqref{eq:qas06}.

We expand
\begin{equation}\label{eq:qas16}
f_k(w)=\sum_{j=0}^\infty f_{jk}(w-\zeta)^j, 
\end{equation}
and obtain
\begin{equation}\label{eq:qas17}
\Phi_k\sim \sum_{j=0}^\infty f_{jk} \Psi_j(\zeta),
\end{equation}
where
\begin{equation}\label{eq:qas18}
\Psi_j(\zeta)=e^{-\frac12\mu\zeta^2}\int_0^\infty e^{-\mu(\frac12 w^2-\zeta w)} (w-\zeta)^j\,dw.
\end{equation}

These functions can be expressed in terms of the complementary error function defined in \eqref{eq:ar01}. 
The first ones are
\begin{equation}
\label{eq:qas19}
\Psi_0(\zeta)=\sqrt{\frac{\pi}{2\mu}}\erfc\left(-\zeta\sqrt{\mu/2}\right),
\quad \Psi_1(\zeta)=\frac{1}{\mu}e^{-\frac12\mu\zeta^2},
\end{equation}
and the other ones follow from integrating by parts in \eqref{eq:qas18}, giving
\begin{equation}\label{eq:qas20}
 \Psi_j(\zeta)=\frac{j-1}{\mu}\Psi_{j-2}(\zeta)+\frac{(-\zeta)^{j-1}}{\mu}e^{-\frac12\mu\zeta^2},\quad j=2,3,4,\ldots.
\end{equation}
By substituting in \eqref{eq:qas18} $w=\zeta+\sqrt{2/\mu}s$ we have
\begin{equation}\label{eq:qas21}
\Psi_j(\zeta)=\left(\frac{2}{\mu}\right)^{(j+1)/2}\int_{-\zeta\sqrt{\mu/2}}^\infty e^{-s^2} s^j\,ds.
\end{equation}
It  follows that $\Psi_j(\zeta)$ can be written in terms of the incomplete gamma functions, see~\eqref{eq:igfs}.

Combining these expansions, we find
\begin{equation}\label{eq:qas22}
Q_{\mu+1}(\mu x,\mu y)\sim\sqrt{\frac{\mu}{2\pi}}
\sum_{j=0}^\infty A_j\Psi_j(\zeta), \quad A_j\sim \sum_{k=0}^\infty \frac{f_{jk}}{\mu^k}.
\end{equation}
We can rearrange this expansion in the form
\begin{equation}\label{eq:qas23}
Q_{\mu+1}(\mu x,\mu y)\sim\sqrt{\frac{\mu}{2\pi}}
\sum_{k=0}^\infty B_k, \quad  B_k=\sum_{j=0}^k \frac{f_{j,k-j}\Psi_{j}(\zeta)}{\mu^{k-j}}.
\end{equation}
The first coefficients  $f_{jk}$ are given in \eqref{eq:comf03}.

By separating the first term in this expansion, and using \eqref{eq:qas19}, we obtain
\begin{equation}\label{eq:qas24}
Q_{\mu+1}(\mu x,\mu y)\sim\tfrac12 \erfc\left(-\zeta\sqrt{\mu/2}\right)
+\sqrt{\frac{\mu}{2\pi}}\sum_{k=1}^\infty B_k.
\end{equation}

If we wish we can find the asymptotic representation for $Q_{\mu}(\mu x,\mu y)$ by using \eqref{RecQ}, which with the present notation can be written as
\begin{equation}\label{eq:qas25}
Q_{\mu}(\mu x,\mu y)=Q_{\mu+1}(\mu x,\mu y)-e^{-\frac12\mu\zeta^2} e^{-\mu\eta(\xi)}I_\mu(\mu \xi).
\end{equation}
This follows from 
\begin{equation}\label{eq:qas26}
\tfrac12\ln(y/x)-x-y+\eta(\xi)=\phi(z_0)-\phi(\xi)=-\tfrac12\zeta^2.
\end{equation}
The result is:
\begin{equation}\label{eq:qas27}
Q_{\mu}(\mu x,\mu y)\sim\tfrac12 \erfc\left(-\zeta\sqrt{\mu/2}\right)+\sqrt{\frac{\mu}{2\pi}}\sum_{k=1}^\infty B_k-
e^{-\frac12\mu\zeta^2} e^{-\mu\eta(\xi)}I_\mu(\mu \xi).
\end{equation}

\subsection{Asymptotic expansion of  \protect{\boldmath{$P_{\mu+1}(\mu x,\mu y)$}}}\label{sec:aspmu}
The computation of the complementary function $P_\mu(x,y)$ is required when  $y<x+\mu$.  To avoid using the relation $P_\mu(x,y)=1-Q_\mu(x,y)$, which  may cause a large relative error, we derive an expansion that can be used for the $P-$function, and that has a similar form and is in fact complementary relation with the expansion in \eqref{eq:qas24}.

Starting from \eqref{eq:asrep05} we have (cf.~\eqref{eq:qas01}) 
\begin{equation}\label{eq:pas01}
P_{\mu+1} (\mu x,\mu y)=\frac{\mu e^{-\mu x}}{(2x)^{\mu+1}}
\int_0^{\xi}z e^{-\mu\phi(z)}e^{-\mu\eta(z)}I_{\mu} (\mu z) \,dz,
\end{equation}
where $\xi$ is defined in \eqref{eq:asrep06} and $\phi(z)$ is the same as in \eqref{eq:qas01}. This function has a saddle point $z_0$ defined in \eqref{eq:qas04}. When $y<x+1$ (in the scaled variables used in \eqref{eq:pas01}), we have $z_0>\xi$.

We use the same mapping as in \eqref{eq:qas06}, now with the conditions that $z=0$ corresponds to $w=+\infty$ and (again) $z=\xi$ to $w=0$. This gives (cf.~\eqref{eq:qas09})
\begin{equation}\label{eq:pas02}
P_{\mu+1}(\mu x,\mu y)=
-\sqrt{\frac{\mu}{2\pi}} e^{-\frac12\mu\zeta^2}
\int_{0}^{\infty} e^{-\mu(\frac12w^2-\zeta w) } f(w)\,dw,
\end{equation}
where $f(w)$ has the same form as in \eqref{eq:qas10}, although the relation between $z$ and $w$ is different; for example $dz/dw<0$. The quantity $\zeta$ is defined in \eqref{eq:qas08} and it is positive in this case, because $y<x+1$.

We can repeat the procedure that we used for the $Q-$function, and the main change is the choice of the square root in $a_1$ that occurs in the expansion in \eqref{eq:comf01}, which has its effect on signs of the coefficients $f_{j,k}$ given in \eqref{eq:comf03}. In this way we obtain the expansion
(cf.~\eqref{eq:qas24})
\begin{equation}\label{eq:pas03}
P_{\mu+1}(\mu x,\mu y)\sim\tfrac12 \erfc\left(\zeta\sqrt{\mu/2}\right)
+\sqrt{\frac{\mu}{2\pi}}\sum_{k=1}^\infty B_k^{*},
\end{equation}
where 
\begin{equation}\label{eq:pas04}
B_k^{*}=\sum_{j=0}^k (-1)^j\frac{f_{j,k-j}\Psi_{j}(-\zeta)}{\mu^{k-j}},
\end{equation}
and $\Psi_{j}(-\zeta)$ can be obtained in the same way as  $\Psi_{j}(\zeta)$ in \eqref{eq:qas19}--\eqref{eq:qas21}. In this representation, the coefficients $f_{j,k}$ are the same as those used for the $Q-$function, and the first few are given in \eqref{eq:comf03}.

From \eqref{eq:qas25}  we obtain
\begin{equation}\label{eq:pas05}
P_{\mu}(\mu x,\mu y)=P_{\mu+1}(\mu x,\mu y)+e^{-\frac12\mu\zeta^2} e^{-\mu\eta(\xi)}I_\mu(\mu \xi).
\end{equation}

\subsection{Where to use the expansions}\label{sec:whereas}

From \eqref{eq:qas21} we see that $\{\Psi_j(\zeta)\}$ is an asymptotic sequence for large $\mu$ and bounded values of $\zeta\sqrt{\mu/2}$. By assuming $-b\le\zeta\sqrt{\mu/2} \le b$, for some $b>0$, we  try to find a domain in the $(x,y)-$plane where we can use the expansions in \eqref{eq:qas24} and \eqref{eq:pas04}. 

For large values of $\mu$ the inequalities can only be satisfied when $\vert\zeta\vert$ is small. In that case we consider the expansion given in \eqref{eq:comf06}, and we use for small values of $\zeta$ the approximation $\zeta\sim -(y-x-1)/\sqrt{(2x+1}$. This gives for $y$ the inequalities
\begin{equation}\label{eq:where01}
x+1-b\sqrt{2/\mu}\sqrt{2x+1}<y<x+1+b\sqrt{2/\mu}\sqrt{2x+1}.
\end{equation}
This is in the scaled variables for $Q_\mu(\mu x,\mu y)$. For the unscaled variables we have
\begin{equation}\label{eq:where02}
x+\mu-b\sqrt{4x+2\mu}<y<x+\mu+b\sqrt{4x+2\mu}.
\end{equation}
From numerical tests we will conclude which values $\mu$ and $b$ can be used for a given set of coefficients in the expansions.

\subsection{Computational aspects of the expansion}\label{sec:compas}
We consider some numerical aspects of the expansion of $Q_{\mu}(\mu x,\mu y)$. First we observe that verifying numerically the parameter domain for applying the asymptotic expansion for $Q_{\mu}(\mu x,\mu y)$ we can use the inhomogeneous recursion given in \eqref{eq:qas25}.

\subsubsection{Expanding $\zeta$}\label{sec:expz}
The value of $\zeta$, see \eqref{eq:qas07} and \eqref{eq:qas08}, vanishes when $z_0=\xi$, that is, when $y=x+1$.
To avoid numerical cancellation we expand as follows. We have
\begin{equation}\label{eq:comf04}
\phi(z_0)=-x-\ln(2x),\quad
\phi(\xi)=y-\sqrt{1+4xy}-\ln\left(\sqrt{1+4xy}-1\right).
\end{equation}
This gives
\begin{equation}\label{eq:comf05}
\tfrac12\zeta^2=\phi(\xi)-\phi(z_0)=x+y-\sqrt{1+4xy}+\ln\frac{1+\sqrt{1+4xy}}{2y}.
\end{equation}
Expanding the right-hand side of \eqref{eq:comf05} in powers at $y=x+1$, we obtain an expansion that can be written in the form
\begin{equation}\label{eq:comf06}
\zeta=-\frac{y-x-1}{\sqrt{2x+1}}\sum_{k=0}^\infty c_k z^k, \quad z=\frac{y-x-1}{(2x+1)^2}.
\end{equation}

The first few coefficients are
\begin{equation}\label{eq:comf07}
\begin{array}{ll}
\dsp{c_{0}=  1,} \\[8pt]
\dsp{c_{1}=  -\tfrac{1}{3}(3x+1),} \\[8pt]
\dsp{c_{2}=  \tfrac{1}{36}(72x^2+42x+7),} \\[8pt]
\dsp{c_{3}=  -\tfrac{1}{540}(2700x^3+2142x^2+657x+73),} \\[8pt]
\dsp{c_{4}=  \tfrac{1}{12960}(15972x+76356x^2+177552x^3+181440x^4+1331).}
\end{array}
\end{equation}

\subsubsection{Computing the coefficients $f_{jk}$}\label{sec:compf}
The computation of the coefficients $f_{jk}$ is rather straightforward, although we need a computer algebra package for obtaining enough coefficients for performing numerical calculations.

First we need the coefficients $a_k$ in the expansion
\begin{equation}\label{eq:comf01}
z=\sum_{k=0}^\infty a_k(w-\zeta)^k, \quad a_0 = z_0.
\end{equation}
These follow from the transformation in \eqref{eq:qas06}. To simplify the notation we introduce $u\in(0,1]$ by writing 
\begin{equation}\label{eq:comf01a}
u=\frac{1}{\sqrt{2x+1}}, \quad x=\frac{1-u^2}{2u^2}.
\end{equation}

Then the first few coefficients $a_k$ are

\begin{equation}\label{eq:comf02}
\begin{array}{ll}
\dsp{a_0=z_0=2\sqrt{x(1+x)}=\frac{1}{u^2}\sqrt{1-u^4},}\\[8pt]
\dsp{a_1= \sqrt{\frac{(2x+1)x}{x+1}}=\frac{1}{u}\sqrt{\frac{1-u^2}{1+u^2}},}\\[8pt]
\dsp{a_2= -a_1\frac{u^3\left(u^2-2\right)}{6\left(1+u^2\right)},}\\[8pt]
\dsp{a_3=-a_1\frac{u^4\left(2u^6+u^4-10u^2+9\right)}{36\left(1+u^2\right)^2},} \\[8pt]
\dsp{a_4=  -a_1\frac{u^5\left(40u^{10}+30u^8-141u^6-158u^4+432u^2-216\right)}{1080\left(1+u^2\right)^3}.}
\end{array}
\end{equation}

The coefficients  $f_{jk}$ are polynomials in $u$ and the first few are

\begin{equation}\label{eq:comf03}
\begin{array}{ll}
\dsp{f_{0, 0}= 1,}\\[8pt]
\dsp{f_{0, 1}= \tfrac{1}{24}u^2\left(3-5u^4\right),}\\[6pt]
\dsp{f_{1, 0}= \tfrac{1}{6}u\left(3+u^2\right),}\\[10pt]
\dsp{f_{0, 2}= \tfrac{1}{1152}u^4\left(81-462u^4+385u^8\right),}\\[6pt]
\dsp{f_{1, 1}=-\tfrac{1}{144}u^3\left(9-21u^2-75u^4+95u^6\right),}\\[6pt]
\dsp{f_{2, 0}= -\tfrac{1}{24}u^2\left(3-5u^4\right),}\\[10pt]
\dsp{f_{0, 3}= \tfrac{1}{414720}u^6\left(30375-369603u^4+765765u^8-425425u^{12}\right),}\\[6pt]
\dsp{f_{1, 2}= -\tfrac{1}{6912}u^5\left(729-1053u^2-9702u^4+11550u^6+12705u^8-14245u^{10}\right),}\\[6pt]
\dsp{f_{2, 1}= \tfrac{1}{576}u^4\left(27-144u^2-402u^4+1440u^6-925u^8\right),}\\[6pt]
\dsp{f_{3, 0}= \tfrac{1}{2160}u^3\left(135-117u^2-675u^4+625u^6\right).}
\end{array}
\end{equation}

\section{A quadrature method}\label{sec:quad}

Numerical quadrature of suitable integral representations can be an important tool
 for evaluating special functions, as explained in \cite[Chapter~5]{Gil:2007:NSF}. In the case of the Marcum functions,
the integrals in \eqref{eq:defQmu} and \eqref{eq:defPmu} give stable integral representations, 
but we prefer a representation in terms of elementary functions. Also, one important 
point for applying the trapezoidal rule  efficiently is the vanishing of the integrand with many (or all) derivatives at the endpoints of integration.

The derivation of the integral representations in terms of elementary functions 
for the Marcum functions is described in detail in \cite{Gil:2012:SAN}. The starting point 
is the contour integral representation (see \cite[Eq.~(2.3)]{Temme:1993:ANA})

\begin{equation}\label{eq:quad01}
Q_\mu(x,y)=\frac{e^{-x-y}}{2\pi i}\int_{\calL_Q}  \frac{e^{x/s+ys}}{1-s}\,\frac{ds}{s^\mu},
\end{equation}
where ${\calL_Q} $ is a vertical line that cuts the real axis in a point $s_0$, with $0<s_0<1$. 

For the complementary function we have

\begin{equation}\label{eq:quad02}
P_\mu(x,y)=\frac{e^{-x-y}}{2\pi i}\int_{\calL_P} \frac{e^{x/s+ys}}{s-1}\,\frac{ds}{s^\mu},
\end{equation}
now with  a vertical line ${\calL_P} $ that cuts the real axis at  a point $s_0$ with $s_0>1$.

Introducing scaled variables $x, y$, we write the representation in \eqref{eq:quad01} in the form
\begin{equation}\label{eq:quad03}
Q_\mu(\mu x,\mu y)=\frac{e^{-\mu(x+y)}}{2\pi i}\int_{\calL_Q}  \frac{e^{\mu\phi(s)}}{1-s}\,ds,
\end{equation}
where
\begin{equation}\label{eq:quad04}
\phi(s)=\frac{x}{s}+ys-\ln s.
\end{equation}

Using saddle point analysis for this expression, we arrive to a final integral representation in the form:

\begin{equation}\label{eq:quad15}
Q_\mu(\mu x,\mu y)=\frac{e^{-\frac12\mu\zeta^2}}{2\pi }\int_{-\pi}^\pi  e^{\mu  \psi(\theta)} f(\theta)\,d\theta, 
\end{equation}

where $\zeta$ is the same quantity as used in \S\ref{sec:asqmu}, see \eqref{eq:qas07}, and

\begin{equation}\label{eq:quad13}
f(\theta)=\frac{\sin\theta\, r^\prime(\theta)+\left(\cos\theta-r(\theta)\right)r(\theta)}{r^2(\theta)-2r\cos\theta r(\theta)+1},
\end{equation}

\begin{equation}\label{eq:quad16}
\psi(\theta)=\cos\theta \rho(\theta,\xi)-\sqrt{1+\xi^2}-\ln\frac{\Frac{\theta}{\sin\theta}+\rho(\theta,\xi)}{1+\sqrt{1+\xi^2}},
\end{equation}

with  $\xi=2\sqrt{xy}$, and $r(\theta)$ and $ \rho(\theta,\xi)$ defined as follows

\begin{equation}\label{eq:quad09}
r(\theta)=\frac{1}{2y}\left(\frac{\theta}{\sin\theta}+\rho(\theta,\xi)\right), 
\quad \rho(\theta,\xi)=\sqrt{\left(\frac{\theta}{\sin\theta}\right)^2+\xi^2}.
\end{equation}

Some extra care is needed when computing the expressions for small values of $\theta$.
Details are given in \cite{Gil:2012:SAN} on how to proceed in this situation
 when computing \eqref{eq:quad13}, \eqref{eq:quad16} and \eqref{eq:quad09}.

Finally, an important point is to determine the domain of applicability of the quadrature method
for computing the Marcum-Q function: 

In \S\ref{sec:whereas} we have explained where we can use the asymptotic expansion derived in \S\ref{sec:asqmu}. 
These are valid around the transition line $y=x+\mu$ for large values of $y,\mu$. See in particular the domain given in \eqref{eq:where02}.
The quadrature method is not valid inside this domain  because the saddle point tends to unity 
(where the pole is located in the representation \eqref{eq:quad01}) as $y\to x+\mu$. So, in the resulting algorithm 
the quadrature method will be used outside the domain specified by \eqref{eq:where02} with a proper value of $b$.
Numerical test leads us to consider $b=1$.  It should be observed that the quadrature method 
does not necessarily need large parameters, although it also performs  quite well in that case.

In \cite{Helstrom:1992:CGM} the trapezoidal rule is used by including the pole at $s=1$ in \eqref{eq:quad03} 
in the function $\phi(s)$. That is, by writing (compare \eqref{eq:quad04})
\begin{equation}\label{eq:where03}
\widetilde{\phi}(s)=\frac{x}{s}+ys-\ln s- \frac{1}{\mu}\ln(1-s).
\end{equation}
In that case the saddle point has to be calculated from a cubic polynomial and
 the  contour follows from  $\Im\widetilde{\phi}(s)=0$ through that saddle point. 
Helstrom used an approximation of this contour by taking a parabola centered at the saddle point.
 The  results in \cite[Table I]{Helstrom:1992:CGM} show correct values at the critical value $y=x+1$,
 also for large values of the parameters, as we will comment later.

\section{Algorithm for the Marcum functions, numerical tests and comparisons}
\label{algorithm}

We propose an algorithm for computing the Marcum functions which
combines different methods of evaluation in different regions: series expansions, integral representations,  
asymptotic expansions, and use of the three-term homogeneous recurrence relation given in \eqref{TTRR}.    

The algorithm uses one or another method based on the expected region of validity of each method supported by
numerical tests of each method, particularly near the boundaries of the domains where we switch methods.

From the numerical tests that we later discuss in detail, the following scheme of computation has been adopted.

We use unscaled variables. Let $\xi=2\sqrt{xy}$ and
\begin{equation}\label{eq:where04}
f_1(x,\mu)=x+\mu -\displaystyle\sqrt{4x+2\mu}, \quad
f_2(x,\mu)=x +\mu + \displaystyle\sqrt{4x+2\mu}.
\end{equation}
Then the scheme is as follows:

\begin{enumerate}
\item{If $x<30$}, then compute the series expansion (\S\ref{sec:userexp}). 
\item{If $\xi>30$ and $\mu^2<2\xi$}, then compute the asymptotic expansion (\S\ref{sec:asrepmus}). 
\item{If $f_1(x,\mu)< y < f_2(x,\mu)$ and $\mu < 135$}, 
then compute the Marcum functions using the recurrence relations \eqref{TTRR}.
\item{If $f_1(x,\mu)< y < f_2(x,\mu)$ and $\mu \ge 135$}, then use the asymptotic
expansion (\S\ref{sec:asrepmul}).
\item{In other case:} compute the integral representation (\S\ref{sec:quad}).
\end{enumerate}

Next we give details on the numerical tests for each method and on the global tests for the complete
algorithm.

\subsection{Testing}

For testing the accuracy of the different methods described in the previous sections, we have used  
the
recurrence relation given in (\ref{RecQ2}). Because, as discussed previously, the $Q$ is neither minimal nor
domininant, the recursion should not be tested in the backward of forward direction; instead,
we write the recurrence in the form

\begin{equation}
\Frac{(x-\mu)Q_{\mu+1} (x,y)+ (y+\mu)Q_{\mu} (x,y)}{xQ_{\mu+2}(x,y)+yQ_{\mu-1} (x,y)}=1.      
\label{errRR}
\end{equation}
and the deviations from $1$ of the left-hand side of \eqref{errRR} (in absolute value) will measure the accuracy of the tested methods. 
We use \eqref{errRR} when $y \ge  x+\mu$ and the same expression but for $P_{\mu}(x,y)$ 
when $y < x +\mu$. For $x>\mu$ all terms in the numerator and the denominator are positive. For the case $x<\mu$
a negative terms appears in the numerator and a more stable test, both for $P$ (when $y < x +\mu$) 
and $Q$ (when $y > x +\mu$) is provided by writing the recurrence as
\begin{equation}
\Frac{(y+\mu)Q_{\mu} (x,y)}{xQ_{\mu+2}(x,y)+(\mu -x)Q_{\mu+1} (x,y)+yQ_{\mu-1} (x,y)}=1.      
\label{errRR2}
\end{equation}

We should note that the tests involve four functions and that the errors in the individual evaluations
of each function turn out to be slightly smaller than the error of the test of the recurrences.
  
We have first checked the implementation of the series expansions given in \eqref{qratio}, which are used for $x<30$.
For computing the ratios of gamma functions appearing in the expressions, we have used the algorithms given in
 \cite{Gil:2012:IGR}.
 As discussed before, the series may present some error degradation for values of the functions
 close to the underflow limit; also, a mild degradation
takes place as large parameters are considered. We have conducted a series of tests using $10^{8}$ random points
in regions such that $(x,y,\mu) \in [0,\,30) \times [0,\,A] \times [1,\,A]$ with increasing $A$. For values of
$P$ greater than $10^{-280}$, our tests show that a relative accuracy better than $10^{-12}$ when 
testing \eqref{errRR} and \eqref{errRR2} can be obtained for $A=200$, while the
 relative accuracies obtained for $A=1000\,,5000\,,10000$
are approximately $3\times 10^{-12}$, $3\times 10^{-11}$ and $5\times 10^{-11}$ respectively. 
For function values smaller than $10^{-280}$
and larger than $10^{-290}$ the relative accuracies are quite uniform for different values of $A$ and around 
$5\times 10^{-11}$.
For function values smaller than $10^{-290}$ the
loss of accuracy becomes more severe. For this reason, we have limited the output of the algorithms to values larger
than $10^{-290}$; and when a smaller value is obtained, we set it to zero.

The  relations \eqref{errRR} and \eqref{errRR} have also been tested for the asymptotic representation for large values 
of $\xi=2\sqrt{xy}$ derived in \S\ref{sec:asrepmus}. 
Our results show that an accuracy better than $6\times 10^{-14}$ 
can be obtained in the parameter region defined by $\xi >30$, $\mu^2 < 2\xi$.
For this case $10^{11}$ random points have been considered in the domain $(x,y,\mu) \in [30,\,10000] \times [0,\,10000]\times [1,\,10000]$ and the values corresponding to random points inside the region of interest and with values not underflowing
have been tested
(close to $8\times 10^{8}$ points). 

The asymptotic expansion for large $\mu$ given in (\ref{eq:qas24}) has been tested 
first  for some of 
 the values of $\mu$, $x$ and $y$ of Table 5.1 in \cite{Temme:1993:ANA} and Table I of \cite{Robertson:1976:CON}.
In these tables, the values of $\mu$ and $y$ are fixed to $\mu=8192$ and $y=1.05\mu$.
The values obtained for $P_{\mu}(x,y)$ and $Q_{\mu}(x,y)$ 
are shown in Table~1 and they are in agreement with those of reference
 \cite{Temme:1993:ANA}. For comparison, we include the values given by MATLAB, and we observe that most of them present
loss of accuracy; we will come back to this point next.
 
\begin{table}
$$
\begin{array}{c|c|c|c}
x/\mu & Q_{\mu}(x,y) & P_{\mu}(x,y) & Q_{\mu}(x,y) \mbox{\, with MATLAB}\\
\hline
0.01 & 1.9845278031193{\rm e-}4  &0.9998015472196881 & 1.9845278031{\bf 062}{\rm e-}4\\
0.02 & 4.138241872117{\rm e-}3& 0.9958617581278824 & 4.13{\bf 4063851051513}{\rm e-}3 \\
0.03 & 0.04000364971081 &   0.9599963502891851 &  0.0{\bf 3999947168921022}    \\
0.04 & 0.191650654805848  &  0.8083493451941514   & 0.1916{\bf 464767819643} \\  
0.05 & 0.498535453743169  &  0.5014645462568305 & 0.49853{\bf 12757146799} \\
0.06 & 0.803520373008492 &   0.1964796269915073 &  0.80352037300{\bf 33010}  \\
0.07 & 0.95565734175388 & 0.04434265824612003 & 0.9556573417{\bf 477112} \\
0.08 & 0.9944737609126645  & 0.005526239087335513 & 0.9944737609{\bf 062473}   \\
0.09 & 0.9996249723836407 &  0.00037502761635937467 & 0.99962497238364{\bf 47} \\
0.1  & 0.9999861372355183 & 0.00001386276448162126  & 0.9999861372355184 \\
\end{array}
$$
{\footnotesize {\bf Table 6.1}. Values for $P_{\mu}(x,y)$ and $Q_{\mu}(x,y)$ obtained by using the asymptotic
expansion of \eqref{eq:qas24}. The values of $\mu$ and $y$ are fixed to $\mu$=8192 and $y=1.05\mu$.  The last column shows the values obtained for
  $Q_{\mu}(x,y)$ with the MATLAB
function {\bf marcumq}; bold font indicates the wrong digits in the MATLAB computations. }
\label{table1}
\end{table}

We have also used the asymptotic expansion of \S\ref{sec:asrep} for the large values of $\mu$ (up to $\mu=10^9$) as shown in
\cite[Table I]{Helstrom:1992:CGM}, and we confirm the shown values of that table.
 These results have been tested with Maple.  
Finally, similarly to the other methods, we have checked the accuracy obtained with the large $\mu$-expansion when
testing the recurrence relation in the parameter region defined
by the relation $x+\mu -\displaystyle\sqrt{4x+2\mu}<y<x +\mu + \displaystyle\sqrt{4x+2\mu}$.  
We consider $10^{10}$ random points for
values of the parameters such that $(\mu,x,y)\in [136,A]\times [0,A]\times [0,A]$ and we analyze the accuracy when
the asymptotic expansion is evaluated. As expected, our results show that the highest errors occur for the smaller values of $\mu$,
and the accuracy tends to decrease as $A$ becomes large, particularly close to the transition line $y=x+\mu$. We get accuracies better than $3\times 10^{-13}$, $2\times 10^{-12}$ and  $10^{-11}$ for $A=200,\,1000,\,10000$ respectively; 
the percentage of points in these regions where asymptotics for large $\mu$ is used is between $0.01\%$ 
($A=200$) and $0.03\%$ ($A=10000$). We have performed an additional test for values of $\mu$ close to $136$ and the accuracies
confirm the previous results and the fact that the accuracy is better as $\mu$ is higher.

Finally, we have tested \eqref{errRR} and \eqref{errRR2} 
when the quadrature method is used in the regions $[0,A]\times[0,A]\times [1,A]$. 
We consider $10^8$ random points for three different values. We obtain an accuracy better than $6\times 10^{-13}$, 
$1.1\times 10^{-12}$ and $10^{-11}$ for $A=200,\,1000,\,10000$ respectively. It is important to observe that more than $95\%$
of the points in these regions corresponds to quadrature, which means that our algorithm uses quadrature most of the time.
However, the rest of methods are crucial in order to cover regions where the quadrature method looses accuracy.

\subsubsection{Global test}

The algorithm has been implemented in the Fortran 90 module {\bf MarcumQ}, which includes the
Fortran 90 routine {\bf marcum} for the computation of $Q_{\mu}(x,y)$ and $P_{\mu}(x,y)$. After the assembly of the
different methods, we have tested that in the regions
$(x,y,\mu) \in [0,A] \times [0,A] \times [1,A]$ we obtain relative accuracies close to $10^{-12}$, $10^{-11}$
and $5\times 10^{-11}$ for $A=200,1000,10000$ respectively when the function values are larger that $10^{-280}$. 
A total of $10^{9}$ random points was considered for each of these tests. For values between $10^{-280}$ and $10^{-290}$ the
accuracy is close to $5\times 10^{-11}$ for the three values of $A$.
 We set all values smaller than $10^{-290}$ to be equal to zero.

\subsection{Comparison against other software}

As far as we know, the only available software for computing Marcum functions are the functions  {\bf marcumq} 
of MATLAB, which  according to MathWorks documentation implements the algorithm given in \cite{Shnidman:1989:GMF}, 
and the function {\bf MarcumQ} of Mathematica. The MATLAB function is a fixed precision algorithm
while the Mathematica function is a variable precision one. It makes sense to compare our algorithm against
MATLAB's, but it is no so clear how to compare against a variable precision algorithm, particularly when 
Mathematica gives no clue about the methods used.

Mathematica computes the function $\widetilde{Q}_{m}(\alpha,\beta)$ of (\ref{alphabeta}), in our notation 
$Q_{\mu}(x,y)$ with $x=\alpha^2 /2$ and $y=\beta^2 /2$; it also computes the complementary function $P$. Real
parameters $m$, $\alpha$ and $\beta$ are accepted. It appears that the algorithm gives correct results if
a sufficiently high accuracy is considered, but that it may fail if the demanded accuracy is not high enough.
We give two examples of this. For instance, computing with $16$ digits accuracy $\widetilde{P}_{1}(40,20)$ with
the command {\bf N[MarcumQ[1,40,0,20],16]} gives  the wrong result $-7.170326438841136\times 10^{-42}$ while if no
accuracy is declared (with {\bf N[MarcumQ[1,40,0,20]]}) Mathematica gives $1.94499\times 10^{-89}$ which is correct;
with $64$ digits accuracy Mathematica still returns a negative value, but the order of magnitude is correct and with $65$
we get three correct digits. A second example is provided by the computation of $\widetilde{Q}_{1}(31,20)$. 
Computing with $16$ digits accuracy $\widetilde{Q}_{1}(31,20)$ with
the command {\bf N[MarcumQ[1,31,20],16]} gives  the wrong result $7.607098470200109\times 10^{20}$ while if no
accuracy is declared (with {\bf N[MarcumQ[1,31,20]]}) Mathematica gives $1$ which is correct; declaring $37$ digits of
accuracy is needed to obtain three correct digits. These examples seem to indicate that Mathematica uses some approximations
suffering from severe cancellations which need to be computed with high accuracy. It is always possible to get
the correct result, but the accuracy needed is uncertain. Apart from this, the Mathematica algorithm appears to be
much slower than our algorithm. For instance, in the region $(x,y,\mu)\in [0,200]\times [0,200]\times [1,200]$
the average time per function evaluation is of the order of $1$ second for computing $\widetilde{Q}$, and slightly larger
for $\widetilde{P}$; in the region $(x,y,\mu)\in [0,20]\times [0,20]\times [1,200]$ the time is of the order of $0.1\,s$ per
evaluation. This has to be compared with the $2\times 10^{-4}\,s$ per evaluation of our algorithm in the region
  $(x,y,\mu)\in [0,200]\times [0,200]\times [1,200]$. We have used Mathematica 8 in our tests; the numerical values
given before can also be checked online at the WolframAlpha site (www.wolframalpha.com).

 The comparison with MATLAB is more natural because it computes the Marcum function with fixed precision arithmetic.
It is important to observe that our method is more general than the function {\bf marcumq} of MATLAB, which can only
be used for integer $\mu$. MATLAB computes $\widetilde{Q}_m (\alpha ,\beta)$, 
but it does not compute the $P-$function; the only way to
obtain values of $P$ from MATLAB is by computing them as $1-Q$, but then small values of $P$ loose accuracy. In terms of speed,
our algorithm appears to be much faster. In average, MATLAB spends around $8\times 10^{-3} s$ per evaluation 
for $(x,y,\mu)\in [0,200]\times [0,200]\times [1,200]$, while our algorithm spends $2\times 10^{-4}\, s$; 
when $(x,y,\mu)\in [0,20]\times [0,20]\times [1,200]$ 
MATLAB spends an average time of $2\times 10^{-3}\, s$ per evaluation and our algorithm $3\times 10^{-5}\, s$.

Apart from being faster, our algorithm is more accurate and reliable than MATLAB's {\bf marcumq} and we have
found ranges of the parameters where our algorithm is accurate while MATLAB gives wrong results.

Some of these accuracy problems occur for large $\mu$, in the region where we consider  the asymptotic
expansion of \eqref{eq:qas24}, as shown in Table \ref{table1}.

Severe problems occur with the MATLAB algorithm near the transition 
line $y=x+\mu$ for $\mu$ large and $x$ small even more, but also for larger values of $y$. 
For example, the minimum (maximum) error found when computing 
$Q_{800}(x,y)$ with $x \in [0.3,\,5]$ and $y \in [806,\,870]$
(taking a step-size in the intervals of $\Delta x=0.2$ and $\Delta y=1$, respectively)
 is $0.87$ ($1.96$). The error is computed as $2\left\|Q_1-Q_2\right\|/(Q_1+Q_2)$,  $Q_1$, $Q_2$ being the values obtained
with MATLAB's {\bf marcumq} and our Fortran 90 module, respectively. 
In other words, none of the 1650 computed 
MATLAB values are correct. 
As a specific numerical example, MATLAB gives $Q_{800}(0.4,810)\simeq 0.0053$ while the correct
value is approximately $0.3632$. 

This type of errors is so clear that it becomes noticeable by just 
plotting function values. 
The plot of the decreasing function $Q_{800}(1,y)$ as a function of $y$ shows different abrupt changes when computed with MATLAB. 
For small $y$ the results are correct, but plotting the function for $y\in [750,850]$ 
we observe a steep jump close to $y=800$ and for $800<y<1100$ all the values are wrong (smaller than the correct values); 
for $y>1100$ a different type of error occurs and the algorithm returns approximately machine-epsilon and for
for $y=2348$ the result jumps to approximately $10^{-300}$ and it jumps again to machine-epsilon when
$2375<y<2384$. Finally, for $y>2384$ the algorithm gives $0$, which is reasonable because the value of $Q$ is already close
to underflow. These anomalies are present for smaller values of $m$. For $m=550$ the first jump is still 
noticeable, and the other phases (machine-epsilon and close to underflow) remain. For smaller values of $m$ the first
jump becomes unnoticeable, but the other (wrong) phases remain for large $y$.

There exist other regions where MATLAB gives no accuracy. For instance, the plot
of the increasing function $Q_{2}(x,200)$ as a function of $x$ shows that the MATLAB function oscillates very rapidly for
$0<x<70$, with most values smaller than machine-epsilon; all these values are wrong.
 Similar errors occur also for large $m$  and even some negative results are obtained. We observe that 
if for $x$ small enough, $Q$ becomes smaller than machine-epsilon, then the MATLAB values are wrong. As $x$ increases, 
the error tends to disappear.

None of these anomalies occur for our algorithm, which is able to compute accurately for parameters smaller than $10000$.
Our main loss of accuracy comes from the use of the series when the functions are close
to the underflow limit, but the results are accurate if they are larger than $10^{-290}$.

\begin{acks}
The authors thank the three anonymous referees for helpful comments and suggestions.
The authors acknowledge financial support from 
{\emph{Ministerio de Econom\'{\i}a y Competitividad}}, projects MTM2009-11686 and MTM2012-34787. 
\end{acks}

\bibliographystyle{acmsmall}
\bibliography{Marcum}

 %
 %
 \end{document}